# Deductive Way of Reasoning about the Internet AS Level Topology[*]

Dávid Szabó[†], Attila Kőrösi[♯], József Bíró[♯] and András Gulyás[‡]

[♯] MTA-BME Future Internet Research Group

[‡] MTA-BME Information Systems Research Group

High Speed Networks Laboratory

Budapest University of Technology and Economics, Hungary

December 9, 2015

**Abstract**

Our current understanding about the AS level topology of the Internet is based on measurements and inductive-type models which set up rules describing the behavior (node and edge dynamics) of the individual ASes and generalize the consequences of these individual actions for the complete AS ecosystem using induction. In this paper we suggest a third, deductive approach in which we have premises for the whole AS system and the consequences of these premises are determined through deductive reasoning. We show that such a deductive approach can give complementary insights into the topological properties of the AS graph. While inductive models can mostly reflect high level statistics (e.g. degree distribution, clustering, diameter), deductive reasoning can identify omnipresent subgraphs and peering likelihood. We also propose a model, called YEAS, incorporating our deductive analytical findings that produces topologies contain both traditional and novel metrics for the AS level Internet.

## 1. Introduction

One simply cannot overestimate the value of knowing more about the topology of the Internet. The last decades have supplied us with thousands of stories where topology-related information about the Internet was directly transformed into more efficient architectures and services, or more appropriate business decisions. The most specific example is clearly Content Delivery Networks (CDN) [1], where global topological peculiarities are highly exploited e.g. in surrogate and cache placement strategies or request routing mechanisms [1] but CDN is just a narrow segment of the whole spectrum. The placement of data centers [2], peer-to-peer networks [3,4], traffic engineering [5], business based AS peering strategies [6], just to mention a few, can clearly benefit from Internet topology related knowledge. The investigation of the AS topology is also a popular topic [7–13] in the network science community, which consolidates researchers from diverse or multidisciplinary research areas. One reason behind this popularity is that compared to other complex networks, active and passive measurements can be executed on the Internet topology, thus we can create Internet "screenshots" easily. Despite the diversity of the contributing research population about the Internet's Autonomous Systems (AS) level topology the way the results are produced seems to be centered around two mainstream approaches.

[*]Paper supported by the High Speed Networks Laboratory

[†]Corresponding author. E-mail: szabod@tmit.bme.hu

One way of getting usable information about the AS topology is simply **measure** it. Today we have historical and contemporary measurement data collected continuously and made publicly available according to various approaches (e.g. using BGP info [14,15], traceroute measurements [16], IXP anatomy [17]). Meanwhile the data stemming from these measurements is the exclusive source of direct information about the AS topology and thus can be treated as the ground truth we can keep ourselves to, the way these measurement systems work is continuously reported to be imperfect and far from optimal [17]. Additionally the collected data reveals only the current state of the network and cannot give usable predictions and clear characterization of the topology forming processes lying in the background.

The other approach is what we can categorize as the **"inductive models"**. Here inductive means that we set up models making premises for the individual network nodes and generalize the results. We can say this is the *de facto* way of getting information about complex networks and basically all known Internet models belong to this category starting from probabilistic random graphs [18], general complex network models e.g. [19], metric space models e.g. [20], fractal models [21], random walk models [18], optimization models [22] but simulation based approaches [23,24] are counted here too. Although the corresponding volume of the literature is considerable still we can identify the following points in almost all cases:

- **Define node dynamics:** This is usually a set of rules regarding how the set of nodes residing in the network varies over time. In the simplest case the set of nodes can be fixed a priori [18] but growth models e.g. [19] where the number of nodes increases over time seems to capture a fundamental aspect of the AS topology.

- **Define edge dynamics:** This is again a rule set controlling how the edges are created between the nodes in the network. The set of rules here can be constructed in a black box fashion [18] where we just want to mimic some high level edge statistics (degree distribution, diameter, clustering etc.) of the AS network, but also can be inspired by processes [19,22] assumed to take place on networks.

- **Analyse and compare with measurement data:** This final and most crucial step where the outcomes of the model are computed, simulated or analytically derived and verified against the available measurement data.

While the above three-step process resulted in a high variety, often precise (as far as the power of measurement data can verify) and usable Internet models, the inductive approach suffers basically from the inability to prove that the processes these models are defined upon, are actually there on the real AS network. For example one cannot really think that preferential attachment in its pure form (where an AS chooses its peers according to their exact nodal degree) is happening in the AS ecosystem. This inability makes the inductive models and their predictions somewhat ambiguous.

*In other words the knowledge we can gain through measurements and inductive models not really focuses on deeper understanding of the networks, as incentives of nodes are unnoticed, and the self-organizing nature of the*

*network formation process remains unrevealed. To cover the self-organizing aspect of networks (and in particular the AS level Internet), in this paper we suggest a third approach which is in line with deductive reasoning thus we can call the produced models **deductive models**.* In the deductive approach we make premises about the whole AS system and analytically prove their exact consequences regarding the topology of the AS level network. Thus the deductive approach will take the following steps:

- **Define premises:** The first and most important part of the deductive approach is defining premises that should capture non-trivial aspects of the system yet permit analytical investigation. Regarding the premises we can make two mistakes: ($i$) we define wrong premises and in this case our results will be irrelevant ($ii$) we define good but weak premises thus we draw valid but meaningless conclusions.

- **Prove the theoretical consequences of the premises:** In this step we construct a model which is compliant with the defined premises and lends itself for analytical investigations. The consequences for the AS topology then should come in the form of theorems with proofs.

- **Compare the results with measurement data:** Mainly serves double-checking purposes if the premises and the analysis are done right.

The above three-step list highlights both the strengths and the weaknesses of the deductive approach. If the premises and the analysis are correct then we gain absolute knowledge about the AS system which provides legitimate predictions until the premises still hold. On the negative side if the premises are wrong, weak or cannot be turned into an analytically tractable model, then we can end up with irrelevant, meaningless or very high level conclusions that cannot be translated into useful messages for Internet practitioners. Finally, we note that we find the deductive approach as somewhat complementary and not supplementary of the inductive one as the two approaches try to tackle the AS network from completely different angles and, as we will see, can provide totally different insights. By understanding the inherent self-organization we have the possibility to predict trends and changes in a complex network that makes us capable to act proactively.

In the rest of this paper we attempt to find a usable set of premises for the Internet AS level topology (Sec. 2)1and construct a model through which the claims stemming from our premises can be formalized and analytically verified (Sec.)3.In Sec. 4we present a topology generator, based on our deductive results, which is able to produce networks that mimic a wide range of features of the AS level Internet. In Sec. 5we conclude by describing the possibilities for future work.

## 2. Deductive analysis of the AS level Internet

In the followings we systematically go through the above three-step process of deductive reasoning regarding the Internet AS topology.

### 2.1. Premises on the AS topology

Over the last four decades the Internet has evolved from a carefully engineered computer network, connecting universities and research institutes in the US, into a complex ecosystem on top of an overwhelming variety of stakeholders all over the world. The network science community emphasizes mostly the resemblance of the AS network to many real world self-organizing networks, which is clearly the case but we argue that this network also has a second face as it apparently exhibits topological peculiarities stemming from technological underpinnings (e.g. the used networking technologies and protocol stacks). It seems that the underlying interdomain routing protocol of the Internet provides a good starting point for finding an analytically tractable set of premises for our deductive reasoning while providing non-trivial insights into the lineament of the AS network's technological face.

Table 1: The simplified BGP best path selection process.

| # | Rule |
|---|---|
| 0. | Valley-free route |
| 1. | Highest local preference |
| 2. | Shortest AS path |
| 3. | Lowest origin type |
| 4. | Lowest MED |
| 5. | eBGP-learned over IBGP-learned |
| 6. | lowest IGP metric to the BGP next-hop |

The interdomain routing policies of all the ASes are expressed through the well defined framework of the Border Gateway Protocol (BGP) [25] [26]. The main responsibility of BGP is to distribute the available forwarding paths between ASes and let them select their preferred paths according to their own special interests. Table 1 recalls a simplified version of the usual steps of the route selection process in BGP from [27] [28]. Here we highlight the vital *valley-free* criteria as rule No. 0, since BGP path selection works over valley-free paths. In this work we pick the first two steps (valley-free policy and local preference) of this process to be our premises as these rules can capture non-trivial aspects of interdomain routing yet permitting analytical tractability.

**Premise 1** (Valley-free routing policy). *In the AS ecosystem the business relationships between ASes can be quite diverse, still we can classify most AS-AS links into basically two major groups [29]: in a* customer-provider *relationship the customer AS pays the provider for forwarding its traffic, while in a* peering *relationship neighboring ASes voluntarily exchange traffic with each other in a settlement-free manner.. The valley-free policy manifests the simple economic principle that the flow of traffic must coincide with the flow of cash. In very short the policy dictates that AS A can use a link to a neighboring AS B to forward the traffic if and only if either the incoming traffic is from a customer or B is a customer of A. Putting it differently valley-free compliant paths comprise arbitrary (may be zero) number of customer-provider links, zero or one peer link and again arbitrary provider-customer links strictly in this order (Fig. 1).*

**Premise 2** (Local preference policy). *The local preference policy is applied on top of valley-free routes meaning that an AS can pick one from the available valley-free routes according to its local interest. Meanwhile these local interests can exhibit high variety the minimalistic rule that customer and peer paths are favored over provider*

---

The authors have published some early and very limited related results. Reference to their work is removed for suiting the double-blind policy.

We omit sibling and backup relationships for simplicity.

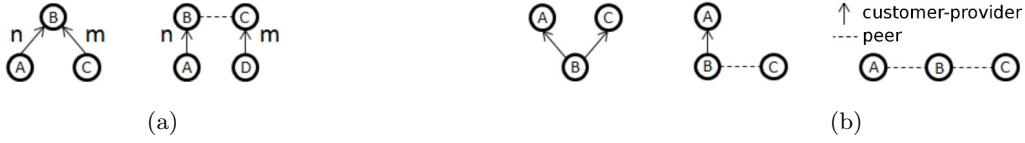

(a)                                                          (b)

Figure 1: Illustration of valid (a) and invalid (b) valley-free path types. A valid path contains $n$ customer-provider, at most 1 peer and $m$ provider-customer link strictly in this order, where $n, m \in \mathbb{N}$. All the other types are invalid paths.

*paths, is contained in basically every local preference setting within the ASes. This is in line with the nature of these routes as customer and peer paths are completely free unlike provider paths in which the provider has to be compensated in some way for the carried transit traffic.*

In what follows we propose a simple game-theoretical model built around Premise 1 and 2 through which their topological consequences can be analyzed.

## 3. Deductive Model and Analysis

Since the AS level topology is formed along the rational business decisions of the individual ASes, game theory is a natural modeling tool of choice. So in the followings we think about the ASes as rational but selfish players whose incentive is to communicate with each other using the policies defined in Premise 1 and 2. More formally, let $\mathcal{P}$ be the set of players (identified as the ASes) with cardinality $N$. According to Premise 1 an edge connecting two players $u, v$ can be of type either $\overline{uv}$ or $\overrightarrow{uv}$, where $\overline{uv}$ denotes a *peer* edge and $\overrightarrow{uv}$ denotes a *customer-provider* edge. The strategy space for player $u \in \mathcal{P}$ is a vector of the preferred edges to other players in the AS network, i.e. the set $S_u = \{(s_{uv})_{v \in \mathcal{P}\setminus\{u\}} : s_{uv} \in \{0, p, r\}\}$ where $|S_u| = 3^{N-1}$ and $p$, $r$ refer to $\overrightarrow{uv}$, $\overline{uv}$ edges, respectively. Easily, player $u$ seeks to contact player $v$ if $s_{uv} \in \{p, r\}$, otherwise $s_{uv} = 0$. We assume simultaneous announcement of the strategies between the the players. Any state of the game is represented by an undirected graph $G(s) = (\mathcal{P}, E(s))$ generated by the strategies of the nodes $s$, where $E(s)$ is given by $E(s) = \{\overrightarrow{uv}|s_{uv} = p \land s_{vu} = 0\} \cup \{\overline{uv}|s_{uv} \in \{r, p\} \land s_{vu} \in \{r, p\}\}$. This settlement of the edges reflects the rational behavior of the ASes as they prefer to create peer edges over customer-provider edges and the instantiation of peer edges requires *bilateral* agreement between the corresponding players while customer-provider edges can be created *unilaterally*.

The goal of the players is to minimize their costs which for a given player $u$ we define as:

$$C_u(s) = \underbrace{\frac{1}{N}\sum_{\forall v \neq u} d_{G(s)}(u, v)}_{\text{communication cost}} + \underbrace{\varphi_p u_p + \varphi_r u_r}_{\text{maintenance cost}}, \quad v \in \mathcal{P} \quad (1)$$

where

$$d_{G(s)}(u, v) = \begin{cases} 0 & \text{if exists a valley free path of which first edge is \textit{peer} or \textit{provider-customer}} \\ 1 & \text{if exists at least one valley free path and the first edge of all of them is \textit{customer-provider}} \\ \infty & \text{if valley free path does not exist} \end{cases} \quad (2)$$

represents the price of communication between $u$ and $v$ over $G(s)$ in compliance with the policies defined in Premise 1 and 2, $\varphi_p$ and $\varphi_r$ are fix maintenance costs of the provider and peer edges and $u_p$ and $u_r$ refer to the number of the $p$ and $r$ edges of $u$ respectively. We note that the cost function in Eq. 1 is intentionally made as simple as possible for two reasons. First we want to concentrate purely on the consequences of our premises thus we avoid incorporating cost elements that can mask them. The second reason is simply analytical tractability. So basically the first sum in Eq. 1 represents the most simple way of capturing our premises and $\varphi_p$ and $\varphi_r$ are introduced for setting up a meaningful game (e.g. without attributing costs to the edges the game would end up in producing full graphs) but can be easily justified as inter AS links clearly have maintenance costs. Also note that we regard provider-customer edges to be financed unilaterally by the customer.

The Nash equilibrium of the game is a state such that no player can further reduce her costs by altering her strategy unilaterally. Since we have a network game we will use the following more natural and slightly tailored equilibrium definition for our case:

**Definition 1** (Pairwise Stable Nash Equilibrium (PNE) [30])**.** *We say $G(s)$ constitutes a pairwise stable Nash equilibrium if (a) Nash equilibrium, (b) $\forall uv \in E(G(s)) : C_u(s) \leq C_u(s') \wedge C_v(s) \leq C_v(s')$, where $s'$ differs from $s$ only in deleting $uv$ edge from $G(s)$, (c) $\forall uv \notin E(G(s)) : C_u(s) \leq C_u(s') \vee C_v(s) \leq C_v(s')$, where $s'$ differs from $s$ only in adding $uv$ edge to $G(s)$ and (d) contains no provider loops.*

Now we are interested in the equilibrium topologies of our game as these topologies will reflect the consequences of our premises. For our claims we need two more definitions.

**Definition 2** (Spider graph (Fig. 2))**.** *A graph is a Spider graph if it consists of:*

1. *a clique $K_r$ (representing the tier-1 ASes) comprising peer edges only*
2. *trees rooted at some subset of $V(K_r)$ having customer-provider edges, such that the provider in the relationship is always closer to the root than the customer*
3. *additional peer edges, such that $\forall \overline{uv}, \overline{uw} \in G(s) : t(v) \cap t(w) = \emptyset$, where $t(x)$ is the set of nodes in the subtree (i.e. the customer cone) of node $x$, including $x$ itself.*

**Definition 3** (Clear-cut Peer Edge (CPE))**.** *An $\overline{uv} \in G(s)$ edge is a clear-cut peer edge if:*

- $\varphi_r < \min\{\frac{|t(u)|}{N}, \frac{|t(v)|}{N}\}$
- $\nexists w \in \mathcal{P} : v \in t(w) \wedge \overline{uw} \in G(s)$.

Our first claim characterizes all meaningful states (i.e. where all the ASes can communicate with each other) of the above game (and thus the AS topology) by identifying an omnipresent subgraph.

**Theorem 1.** *Every meaningful outcome of the game i.e $\sum C_u \neq \infty$ contains the Spider graph as a spanning subgraph.*

*Proof.* The subgraph of the customer-provider edges is a spanning DAG, as provider loops are not allowed. For having $\sum C_u \neq \infty$ the sinks of this DAG has to be connected by peer edges in pairs. Hence the set of the sinks correspond to the $K_r$ clique of the Spider graph.

---

This requirement is fully in line with the Gao-Rexford conditions [31] ensuring BGP stability.

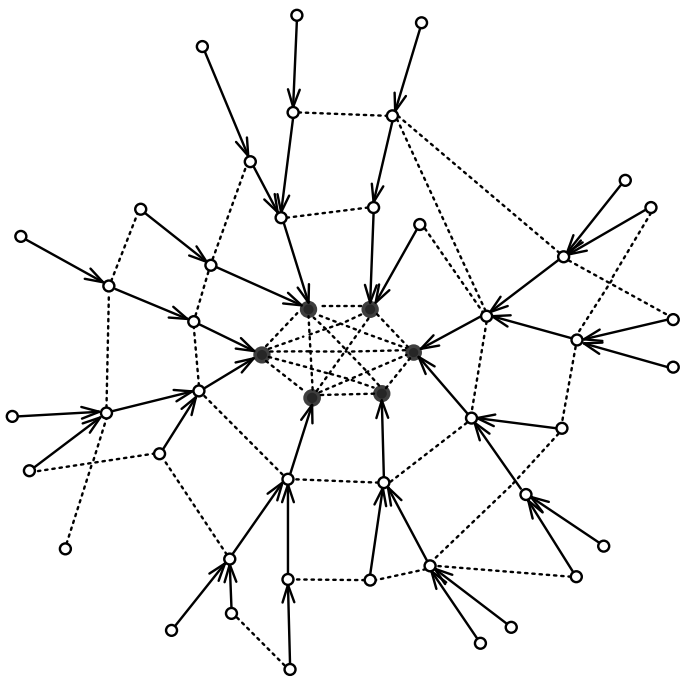

Figure 2: An example of the spider graph, the dashed and directed edges are the peer and customer-provider edges respectively and the black nodes are the ASes of the clique $\mathcal{K}$ i.e. the tier-1 ASes.

Obviously each AS has a directed customer-provider path to some ASes of $K_r$. So one spanning forest of the DAG and the $K_r$ clique is a proper spanning Spider graph in the original graph. $\square$

Using Theorem 1 we can characterize the pairwise stable equilibria of the game.

**Theorem 2.** *Every pairwise stable equilibrium of the game is the Spider graph.*

*Proof.* According to Theorem 1 any pairwise stable equilibrium contains the Spider graph as a spanning subgraph. Easily it contains a $K_r$ clique in which ASes do not have customer-provider edges. If there are any extra customer-provider edges, then there must be an AS which has at least two customer-provider links. Since the additional customer-provider edge does not reduce the communication cost, but enlarges the maintenance cost, such outcome cannot be a Nash equilibrium.

If the subgraph of the customer-provider edges is a forest, then in it if exists two nodes $v$ and $w$ such that $t(v) \cap t(w) \neq \emptyset$, then $v \in t(w)$ or $w \in t(v)$. Hence, if there is a peer edge $\overline{uv}$ such that there exists a node $w$: $\overline{uw} \in E(G)$ and $t(w) \cap t(v) \neq \emptyset$, then $v \in t(w)$ or $w \in t(v)$. Let $w \in t(v)$, so $u$ can reduce its cost removing $\overline{uw}$, which contradicts the definition of the Nash equilibrium. $\square$

The following theorem gives a high-level insight into the placement of the peer edges.

**Theorem 3.** *If $G(s)$ constitutes a PNE then each peer edge is a CPE or part of $K_r$.*

*Proof.* We prove this indirectly. If there exists a peer edge out of $K_r$ which is not CPE then either $(i)$ $\varphi_r \not< \min\{\frac{|t(u)|}{N}, \frac{|t(v)|}{N}\}$ or $(ii)$ $\exists w \in V(G(s)) : v \in t(w) \wedge \bar{u}\bar{w} \in G(s)$. For $(i)$ it is easy to see that at least for one AS it is worth to delete the edge. For $(ii)$ it's trivial that for $w$ is worth to delete $\bar{u}\bar{w}$. In both cases we appear to a contradiction. □

Finally our theorems lead to the following three corollaries.

**Corollary 1.** *In a PNE a peer edge appears only if it is in $K_r$ or its both endpoint ASes has sizable customer cones.*

**Corollary 2.** *For PNEs there exists an upper bound for the size of the customer cones of the ASes in $K_r$, or more formally $PNE \implies \max_{u \in V(K_r)} t(u) \leq N(\varphi_p - \varphi_r(|V(K_r)| - 1) + 1)$.*

*Proof.* The cost of a node $u \in V(K_r)$ is $\varphi_r(|V(K_r)| - 1)$. However, if $u$ leaves $K_r$ and creates only one customer-provider edge to another node in $K_r$, its cost would change to $\frac{N-t(u)}{N} + \varphi_p$. Hence in PNE

$$\varphi_r(|V(K_r)| - 1) \leq \frac{N - t(u)}{N} + \varphi_p, \forall u \in V(K_r), \tag{3}$$

and thus

$$\max_{u \in V(K_r)} t(u) \leq N(\varphi_p - \varphi_r(|V(K_r)| - 1) + 1) \tag{4}$$

□

**Corollary 3.** *In the case of PNE there exists an upper bound for the size of $K_r$ independent from $N$, i.e. $PNE \implies |V(K_r)| \leq \frac{\varphi_p + \varphi_r + 1 + \sqrt{(\varphi_p + \varphi_r + 1)^2 - 4\varphi_r}}{2\varphi_r}$*

*Proof.* According to Corollary 2

$$\max_{u \in V(K_r)} t(u) \leq N(\varphi_p - \varphi_r(|V(K_r)| - 1) + 1), \tag{5}$$

and obviously

$$\frac{N}{|V(K_r)|} = \text{avg}_{u \in V(K_r)} t(u) \leq \max_{u \in V(K_r)} t(u), \tag{6}$$

hence

$$\frac{N}{|V(K_r)|} \leq N(\varphi_p - \varphi_r(|V(K_r)| - 1) + 1). \tag{7}$$

Dividing by $N$ and rearranging the inequality we get:

$$0 \leq -\varphi_r |V(K_r)|^2 + (\varphi_p + \varphi_r + 1)|V(K_r)| - 1, \tag{8}$$

implying

$$|V(K_r)| \leq \frac{\varphi_p + \varphi_r + 1 + \sqrt{(\varphi_p + \varphi_r + 1)^2 - 4\varphi_r}}{2\varphi_r} \tag{9}$$

□

The above theorems deliver the following high-level sketch of the AS topology as a main intuitive message: $(i)$ it is a Spider-like graph with a clique (of tier-1 ASes) in the center and trees routed in the nodes of the clique, $(ii)$ the peer edges appear more likely between ASes having sizable customer cones, $(iii)$ the size of the clique is constrained by the maintenance cost of peer and customer-provider relationships and $(iv)$ the largest customer cone size in the nodes of the clique is also driven by these maintenance costs.

## 3.1. Discussion and double-checking against measurement data

For validating our analytical results we used the AS Relationships dataset of May 2012, provided by CAIDA [14]. Although this dataset received some criticism over the last years, at this moment no other sources of data are available containing more accurate tracing of the peer and customer-provider edges at the AS level.

This dataset contains AS-AS relationships for 41203 ASes with 57158 peer and 83374 customer-provider edges, thus lets us build a labeled AS graph. Regarding Theorem 1 and 2 we investigated the existence of the Spider graph in two steps. First we followed the customer-provider relationships in a top-down manner proceeding from the top tier-1 clique and kept all the nodes we could reach, this way we get a 92.5% node coverage which properly validates that the AS graph meets the first two properties (clique inside and trees rooted on the nodes of the clique) of Spider graphs. Secondly, we examined how typical for an AS $C$ with peering neighbors $A$ and $B$ that $t(A) \cap t(B) = \emptyset$. In other words we calculated how typical is that the customer cones of the peers of an AS are overlapping (this is the direct checking of the third property of Spider graphs see Definition 2). For this we randomly sampled the measured AS graph by choosing 500000 $(A, B)$ node pairs for which $\overline{CA}, \overline{CB}$ exists. In each sample we drew AS $C$ according to a degree-weighted probability function and then we picked the peering neighbors with uniform normal distribution. Our results confirmed that more than 75% of the pairs (Fig. 3) have zero overlapping and in other cases the ratio of overlapping vanishes very quickly. These results readily support our claim that the AS level Internet topology is a Spider-like graph.

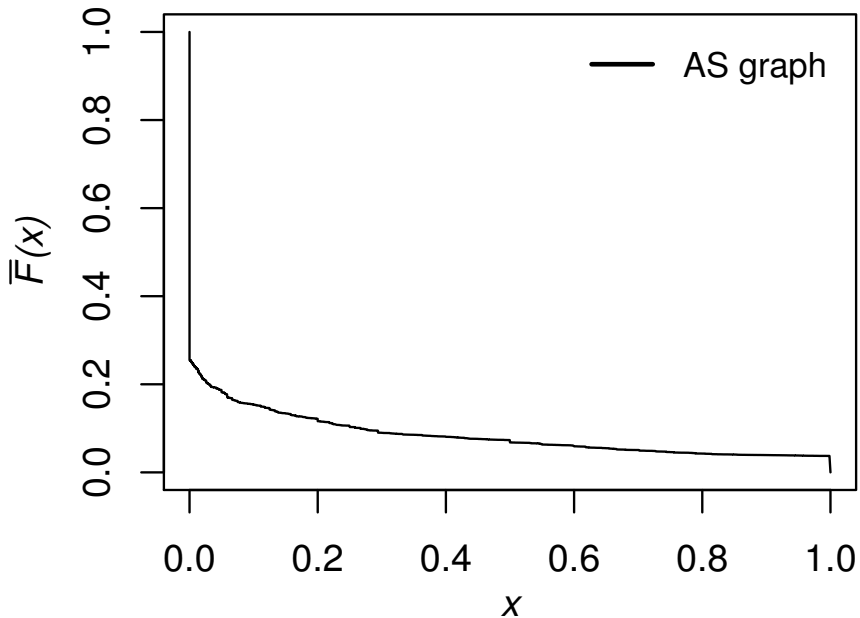

Figure 3: CCDF for coverage overlapping of peer edges of an AS defined as $x = \frac{t(A) \cap t(B)}{min\{t(A), t(B)\}}$

After that as a next step, we measured the peering likelihood between two ASes as a function of the minimum of their customer cone sizes. The AS graph dataset of Fig. 4 shows the empirical probability that two ASes with a given minimum customer cone size $(\min(t(A), t(B)))$ are in a peering relationship. The dataset supports that the peering likelihood is in high correlation with the customer cone sizes of the ASes in the peering relationship.

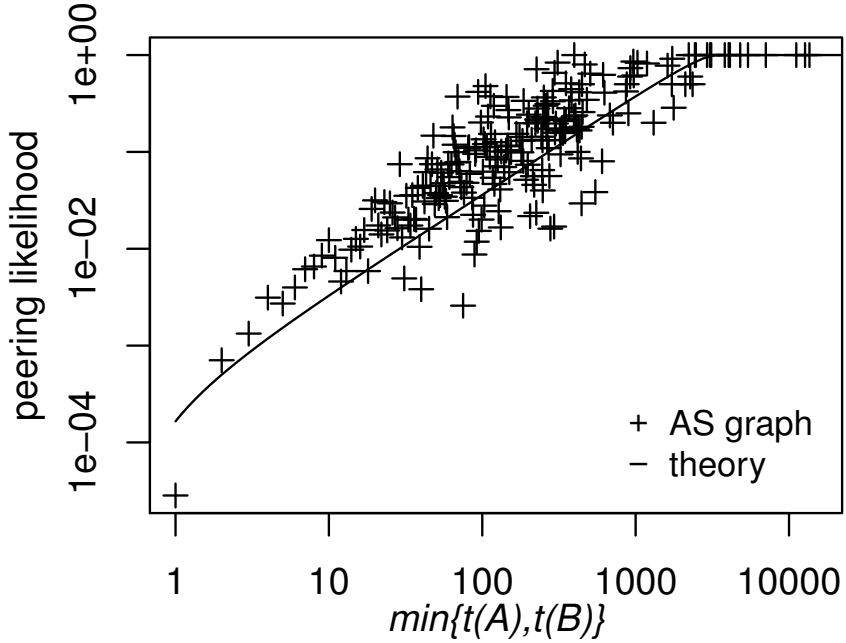

Figure 4: Peering likelihood between ASes as the function of their customer cone size.

Finally, we present a short argument illustrating our deductive predictions on the maximum customer cone size and the max size of the tier-1 clique. For doing this we used historical AS datasets from CAIDA. Based on the number of customer-provider and peering relationships we have estimated $\varphi_p = \frac{Nc_1}{\#\text{of c-p edges}}$ and $\varphi_r = \frac{Nc_2}{\#\text{of peer edges}}$ with $c_1 = 1.1$ and $c_2 = 0.05$. Using these values we have computed the results of our corresponding theorems and measured the max-conesize and tier-1 clique size as a function of time in the CAIDA datasets. Fig. 5 shows that our rough estimation about the maximal customer cone size in the AS level Internet approximates the measured one based on CAIDA snapshots in a reasonable extent. Fig. 6 shows the prediction of our model regarding the size of the tier-1 clique. Although our simple formulae forecast a more increasing trend, the order of magnitudes are quite the same in both cases.

As a discussion we first call the reader to notice the complementary nature of the deductive findings as opposed to the existing inductive models. While the existing inductive models concentrate on degree distribution, clustering, diameter etc. the deductive reasoning give hints about spanning subgraphs, peering likelihood and constraints on the size of different parts of the network. We also recall that our deductive model is extremely simple and squeezes all maintenance cost related quantities into two constants $(\varphi_r, \varphi_p)$. In the light of this simplicity it is remarkable that the model gives practically usable predictions regarding the size of the tier-1 clique and the maximal customer cone of an AS.

One may argue that the results coming out of our deductive analysis are somewhat weak and don't say too much about the AS network. Such criticism may seem to be all right at first, but we find to be important and interesting in itself that the found topological peculiarities (summarized in Theorem 1,2,3 and Corollary 1,2,3) are direct consequences of the used BGP policies and thus will be present on the AS topology as far as these policies are at use. We believe that showing this causality contributes to our very limited amount of information about the Internet AS level topology. Finally we note that more powerful premises can lead to more precise

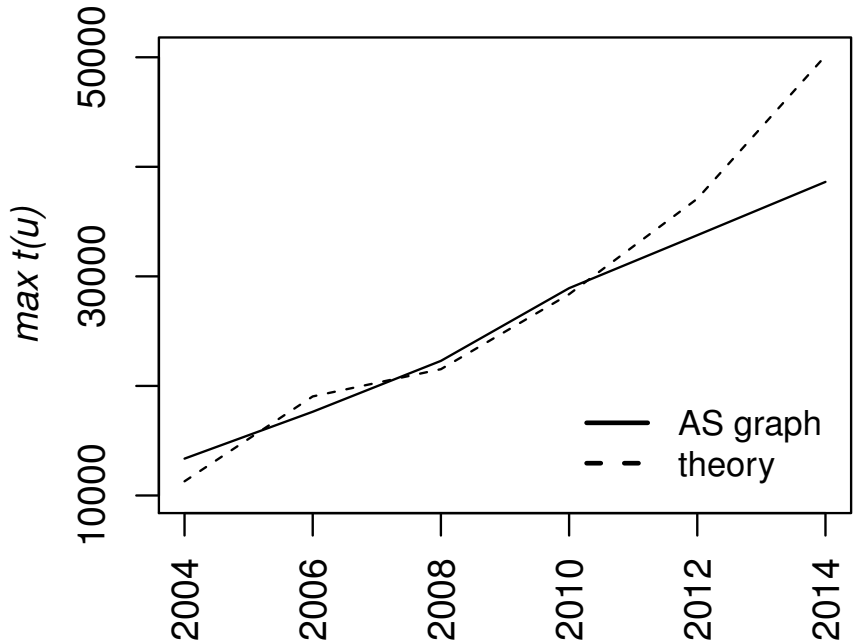

Figure 5: Comparing our upper bound for *max t(u)* based on Corollary 2 with the AS graph over time.

deductive topology characterization in future works.

## 4. A Deductive Inspired AS Level Internet Model (YEAS)

Our deductive analysis provided us some fresh insights into the AS level topology which, at least according to our best knowledge, are not incorporated in existing AS topology models to a proper extent. While the deductive results give useful hints about the general structure of the AS graph, the high-level nature of these hints is a clear obstacle to convert them into practically usable synthetic AS topologies having at least comparable features with respect to the inductive models. So in what follows we build a generative AS topology model called YEAS, but besides recovering the usual features of inductive models (e.g. power-law degree distribution, large clustering, small diameter etc.) we implicitly encode the outcome of our deductive analysis into the node and edge dynamics. Thus at the end of the day we require YEAS to produce Spider-like graphs having correct edge labeling, realistic tier-1 clique size and realistic placement of the peer edges. The framework of YEAS is based on the recently advocated hyperbolic space models presented in [20]. What we basically do is dressing a very simple hyperbolic model with our deductive findings.

### 4.1. Node layout

We distribute the nodes (still representing the ASes) quasi-uniformly on the surface of a hyperbolic disk of radius $R$. This is done by assigning polar coordinates to each node as:

$$r = (1/\alpha)\operatorname{acosh}\left(1 + [\cosh(\alpha R) - 1]\,U_1\right) \tag{10}$$

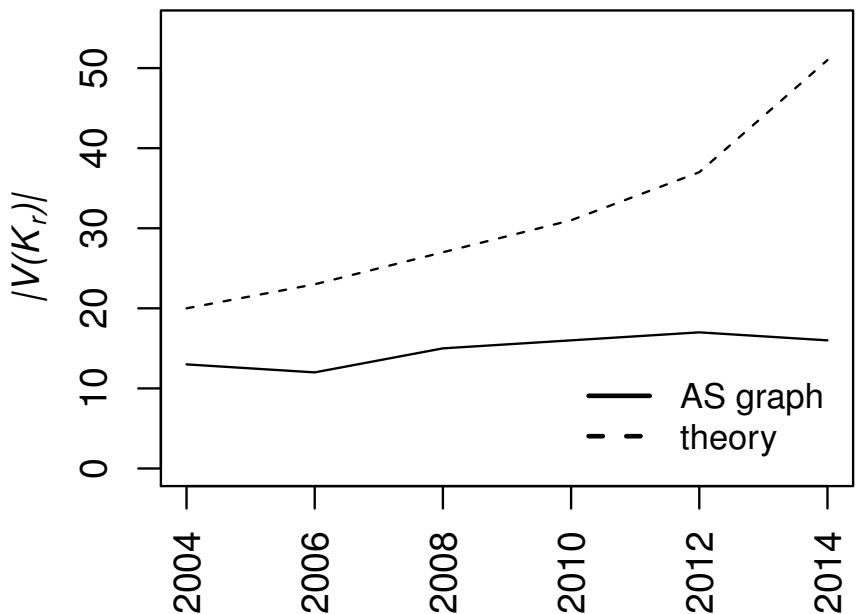

Figure 6: Comparing our upper bound for $|V(K_r)|$ based on Corollary 3 with the AS graph over time.

$$\phi = 2\pi U_2 \tag{11}$$

where $U_1$ and $U_2$ are random independent variables distributed uniformly in the interval $(0,1)$ and $\alpha$ is a parameter controlling the heterogeneity of the layout.

### 4.2. Edge creation

For initialization take node $u$ with the lowest radius and initialize a set $\mathcal{K} = \{u\}$. In the first phase take nodes $w$ one by one in an increasing order of their radii $r_w$ and connect them to the others according to the following simple rule: if

$$Q \sum_{v \in \mathcal{K}} l(r_w, \phi_w, r_v, \phi_v) < \min_{v | r_v \leq r_w} l(r_w, \phi_w, r_v, \phi_v), \tag{12}$$

then connect $w$ to all nodes in $\mathcal{K}$ with peer edges and add $w$ to $\mathcal{K}$, otherwise connect $w$ to node $\mathrm{argmin}_{v | r_v \leq r_w} l(r_w, \phi_w, r_v, \phi_v)$ with a customer-provider edge. The constant $Q$ is a tunable model parameter controlling the size of $\mathcal{K}$ and

$$l(r_u, \phi_u, r_v, \phi_v) = \mathrm{ach}(\mathrm{ch}r_u \mathrm{ch}r_v - \mathrm{sh}r_u \mathrm{sh}r_v \cos(\phi_u - \phi_v)). \tag{13}$$

In the second phase every node $u \notin \mathcal{K}$ connects to a node $v$ with peer edge if $\nexists \overrightarrow{uv} \wedge l(u,v) < \beta R$, where $\beta$ is a parameter in the interval $(0,1)$ for tuning the peering willingness. The pseudocode of the complete process is shown in A for the sake of reproducibility.

---

In YEAS this set represents the clique of tier-1 ASes.

### 4.3. Analysis and comparison with measurement data

First we show that YEAS can readily *recover power law degree distribution and high clustering coefficient*, which can be observed in the real AS topology. For doing this we show, that our model generates all edges $(u,v)$ for which $l(r_u, \phi_u, r_v, \phi_v) < R$ but contains edges $(u,v)$ for which $l(r_u, \phi_u, r_v, \phi_v) > R$ with negligible probability. The first statement trivially follows from the edge creation process itself. The second one is the direct consequence of equation (16), that is if the distance between two points $l$ is greater than $R$, the probability having an edge between these two points is smaller than $e^{-\delta e^{\frac{R}{2}}}$. (See the derivation of (16) later in this subsection).

By simply but arguably ignoring (at least from the point of degree distribution and clustering) the negligible number of edges of length larger than $R$, we end up with a model readily analyzed in [20]. Nevertheless we recall the most important claims for making our argument self-contained. The expected degree of a node with coordinates $r, \phi$ is the number of expected points within its $R$−distance vicinity. Equivalently, this coincides the expected number of points falling inside the intersection of the original $R$−disk and the disk with radius $R$ and center $r, \phi$. In case of $0.5 < \alpha \leq 1$ the degree of a node with radial coordinate $r$ decays exponentially in the function of $r$ (approximately independent from $\alpha$), $\bar{k}(r) \sim e^{-\frac{r}{2}}$, while the node density exponentially increases, $\rho(r) \sim e^{\alpha r}$. The combination of these two exponentials yields a power law degree distribution $P(k) \sim k^{-2\alpha-1}$, and complement cumulative degree distribution $\bar{F}(k) \sim k^{-2\alpha}$ [32] [33]. It can be rigorously shown that there exists a constant lower bound on the global clustering coefficient in hyperbolic random graphs, which confirms the high clustering claimed in such networks [34].

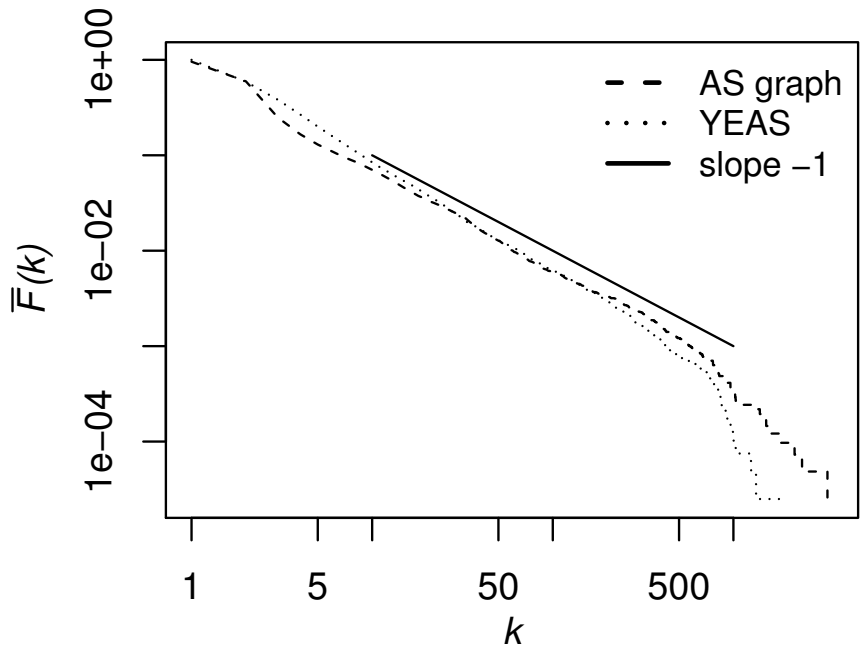

Figure 7: CCDF of degrees in the real AS graph and in the YEAS topology.

Fig. 7 shows the cumulative degree distribution of the real AS graph compared to the degree distribution of YEAS with the setting $N = 40000$, $Q = 5$, $\alpha = 0.55$, $\beta = 0.7$ and $R = 18.5$ (we use this setting for all

---

For reasonable value of $N = 40000$, $R = 17.9$, this probability is 0.00135

the simulations from now on). The measured AS graph contains 41203 nodes, so we generated a similar sized topology. The clustering coefficient for the AS graph and for YEAS are both high 0.38 and 0.69 respectively. Table 2 provides additional metrics for comparison.

Table 2: Comparison of a YEAS generated topology and CAIDA topology for basic metrics.

| Network | Nodes | Edges | C. coef. | Avg. dist. |
|---|---|---|---|---|
| CAIDA top. | 41203 | 116930 | 0.38 | 3.81 |
| Our model | 40000 | 115309 | 0.69 | 4.07 |

| Network | Avg. degree | Diameter | Max. cluster | # Tier-1 |
|---|---|---|---|---|
| CAIDA top. | 5.67 | 14 | 39327 | 16 |
| Our model | 5.76 | 12 | 40000 | 16 |

Now turn to quantities almost never tackled by the existing inductive models. The first one will be the *expected customer cone size of the nodes and the cone size distribution of the whole network*. Building upon these results we will be able to quantify the peering probability of the nodes not residing in $\mathcal{K}$. For analyzing the average customer cone sizes we temporally omit the peer edges generated by the model as these do not effect the customer sizes. For simplicity we conduct the analysis for $C = \infty$ but later on we show that these results coincides with simulation results.

Let $p(s, \phi, r)$ denote the probability by which node $u$ with radial coordinate $s$ and angle $\phi$ establishes a provider link to node $v$ with radial coordinate $r$ and angle 0 provided that $s > r$. Recall that node $u$ establishes a customer-provider link to node $v$ if and only if $s > r$ and node $v$ is the closest to node $u$. The equivalent geometrical meaning of this condition in the generation model is that none of the other $N - 2$ points fall within the intersection of the circle with radius $s$ (with the same center as of the $R$−disk) and the $(s, \phi)$−centered circle with radius $l(s, \phi, r, 0)$. With using elementary hyperbolic geometrical properties, the area of the intersection of these circles can be well approximated (if $l(s, \phi, r, 0)$ is not very close to 0) as

$$A_{\text{intsec}} \approx 4\mathrm{e}^{\frac{l(s,\phi,r,0)}{2}} . \tag{14}$$

Now the probability that none of the other $N - 2$ points fall within this intersection area can be formulated as

$$\left(1 - \frac{A_{\text{intsec}}}{A_{R-disk}}\right)^{N-2} \approx \mathrm{e}^{-\delta A_{\text{intsec}}} \tag{15}$$

where $\delta = \frac{N}{A_{R-disk}}$ is defined as the average node density. The approximation of $p(s, \phi, r)$ is now resulted as

$$p(s, \phi, r) = \mathrm{e}^{-\delta 4 \mathrm{e}^{\frac{l(s,\phi,r,0)}{2}}} . \tag{16}$$

The expected customer cone size as a function of $r$ $\bar{T}(r), r = 0, \ldots, R$ fulfills the following integral equation.

$$\bar{T}(r) = 1 + N \int_{s=r}^{R} \int_{\phi=0}^{2\pi} \bar{T}(s) p(s, \phi, r) \rho(s) \mathrm{d}\phi \mathrm{d}s . \tag{17}$$

---

We can assume that the angle coordinate of node $v$ is 0 without loss of generality.

where $\rho(s) = \frac{\sinh(u)}{2\pi(\cosh(R)-1)}$ is the node density function. The intuitive explanation of Eq. 17 is the following. The customer cone of a node $v$ with radical coordinate $r$ consists of itself and all other nodes' cones with larger radial coordinate $s > r$ and any angle coordinate $\phi$, which are connected to node $v$ by probability $p(s, \phi, r)$. For reformulating this equation the following approximations are used: $\rho(s) \approx \frac{1}{2\pi}e^{s-R}$, $\delta = \frac{N}{A_{R-disk}} \approx \frac{N}{\pi e^R}$. By applying these the integral equation becomes

$$\bar{T}(r) = 1 + \frac{\delta}{2}\int_{s=r}^{R} \bar{T}(u)\left(\int_{\phi=0}^{2\pi} e^{-4\delta e^{\frac{l(s,\phi,r,0)}{2}}}d\phi\right) e^s ds \ . \tag{18}$$

The inner angle integral can be well approximated as $\frac{1}{\delta}e^{\frac{-s-r}{2}}$. This provides the following (approximate) form of the integral equation:

$$\bar{T}(r) = 1 + \frac{1}{2}\int_{s=r}^{R} \bar{T}(s)e^{\frac{s-r}{2}}ds \ . \tag{19}$$

The solution of this integral equation gives the function $\bar{T}(r)$. Unfortunately it can not be solved analytically, however, the solution can be readily characterized as an exponential function. The detailed investigation of the numerical solution confirms this intuition, for a wide range of radial coordinates $r$, the function $\bar{T}(r)$ is approximately proportional to $e^{-r}$.

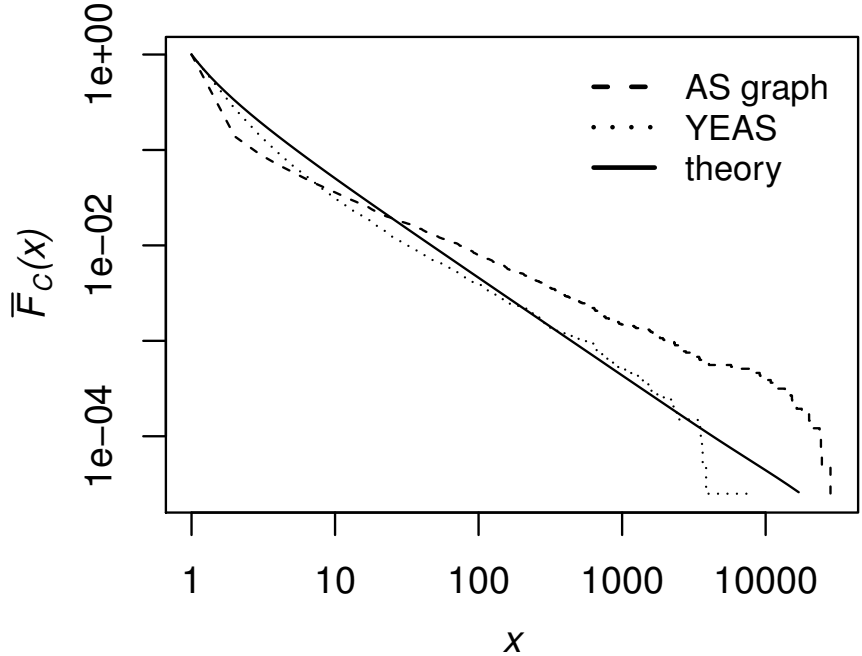

Figure 8: CCDF of customer cone sizes in the real AS graph, theory and in the YEAS topology.

Now we can analyze the complement cumulative distribution of cone sizes $\bar{F}_T(x) = P(T > x)$. The CCDF of the cone sizes is approximately a power-law with exponent $-1$ provided the expected cone size given $r$ is proportional to $e^{-r}$, that is $P(T > x) \approx x^{-1}$ . Fig. 8 readily supports this results as the theoretical result goes hand in hand with the outcome of our simulations. Comparing the real AS topology we detect slightly smaller customer cone sizes produced by YEAS. This is definitely the lack of multihoming in the current version of our

---

In case of sparse networks, the conditional distribution of $T(r)$ is Poissonian with mean $\bar{T}(r)$, $P(T(r) = x) = \frac{\bar{T}(r)^x}{x!}e^{-x}$. Deconditioning this w.r.t. $r$ results a distribution approximately proportional to $x^{-2}$, therefore the CCDF of $T$ will be approximately proportional to $x^{-1}$ .

model. In the AS graph there are many ASes having multiple providers in order to increase reliability, thus many AS contributes to the cone size of multiple ASes. Nevertheless the tendency of the cone size distribution is correctly recovered by YEAS, although the exponent is not exactly the same.

Finally we can turn to analyzing the *peering likelihood* $P_{peering}$ of two nodes having expected customer cone sizes $\bar{T}_1, \bar{T}_2$. More explicitly, we determine the peering probability in the function of $\min(\bar{T}_1, \bar{T}_2)$. For this, first the peering probability of two nodes with radial coordinate $r_1$ and $r_2$ in the function of $\max(r_1, r_2)$ is calculated, then the function $r(\bar{T})$ (the inverse function of $\bar{T}(r)$) is applied. Without loss of generality, assume that $r_1 < r_2$. Given $r_2$, the nodes with smaller radial coordinates $r_1 < r_2$ lies within the circle with radius $r_2$ and center 0. Clearly, among these nodes those have peer edges to node $r_2$ which lie in the intersection of this disk and the $r_2$ centered $R$−radius disk. Therefore, due to the uniform distribution of nodes the peering probability is the ratio of this intersection area and the area of the 0-centered $r_2$ radius disk. Evidently, the peering probability is 1, if $r_2 < \frac{R}{2}$, because in this case the 0-centered $r_2$ disk is fully contained by the $r_2$−centered $R$−radius disk. If $r_2 > \frac{R}{2}$ it can be shown by elementary hyperbolic geometry that

$$P_{peering}(r_2, R) \approx \frac{\arccos\left(\frac{\cosh^2(r_2) - \cosh(R)}{\sinh^2(r_2)}\right)}{\pi}$$
$$+ \frac{\exp(R)\arccos\left(\frac{\cosh(r_2)\cosh(R) - \cosh(r_2)}{\sinh(r_2)\sinh(R)}\right)}{\pi \exp(r_2)} \ . \tag{20}$$

The more detailed analysis of this approximation discloses that it is well approximately proportional to $\mathrm{e}^{-r}$. From this a simple enough approximation can be obtained as

$$P_{peering}(r_2, R) \approx \begin{cases} 1 & \text{if } r_2 < \frac{R}{2} \\ \mathrm{e}^{\frac{R}{2}}\mathrm{e}^{-r_2} & \text{if } r_2 \geq \frac{R}{2} \ . \end{cases} \tag{21}$$

It apparently follows that

$$P_{peering}(\bar{T}_2, R) \approx \begin{cases} 1 & \text{if } \bar{T}_2 > \bar{T}(\frac{R}{2}) \\ \frac{1}{\bar{T}(\frac{R}{2})}\bar{T}_2 & \text{if } \bar{T}_2 \leq \bar{T}(\frac{R}{2}) \ . \end{cases} \tag{22}$$

This means that the likelihood of peer edges of an AS having a customer cone size to other ASes having larger customer cone sizes is proportional to its cone size, and this likelihood tends to be 1, if the cone size falls below a certain limit. This characteristic property is also confirmed by our simulations shown in Fig. 9 and coincide with results measured on the real AS topology.

The above theoretical results show that YEAS generates realistic complex networks with proper degree distribution, clustering and diameter, yet incorporating our deductive findings as the synthesized topologies are Spider-like (trivially follows from the generation process), with tunable tier-1 clique (through the $Q$ parameter) and realistic peering likelihood.

## 5. Conclusion and Future work

In this paper we have proposed a deductive approach for revealing network dynamics on the Internet's AS level topology formation. We believe this approach allows us to make some reasonable predictions about the structural changes of complex networks and facilitates the design of proactive algorithms. The first step for this is to find the correct premises, which is not trivial, but the findings of other complex networks can also be used as a starting point and even can reinforce each other. For example it is possible that valley-free information flow is not just an Internet specific feature. At the highest level it says: "a minor node cannot serve the demand between two major nodes", which can be used also in neural or in social networks (after careful consideration, of course).

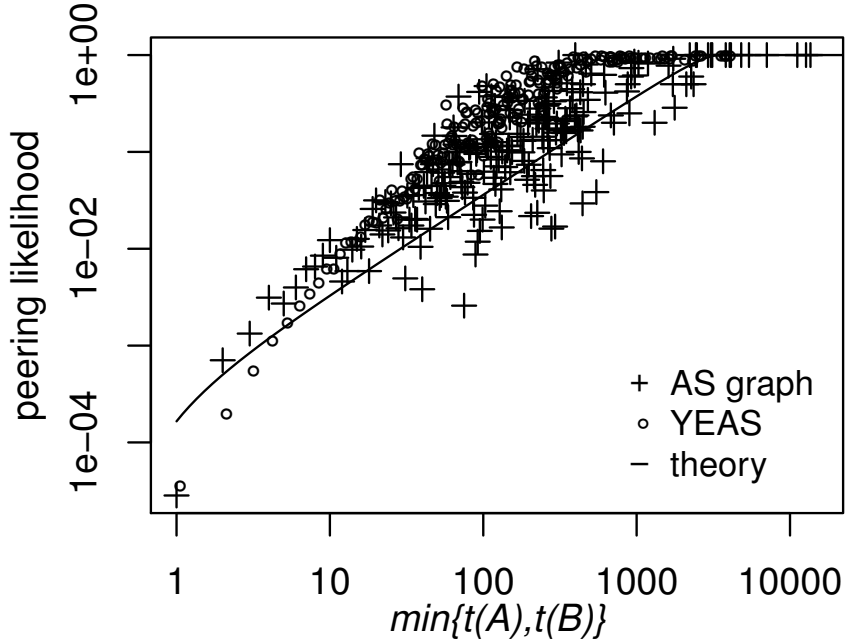

Figure 9: Peering likelihood between ASes as the function of their customer cone size (here we exteneded Fig. 4 by adding results about the YEAS generated topology).

So in accordance to this approach we made premises about the AS level Internet that have provided us the possibility of theoretical analysis and let us improving our interpretation of the AS system as provide insights complementary to the existing inductive models. One can question which other directions we should look for more premises that can shed more light on the AS level topology. Our first suggestion would be to dig more deeply into the interdomain route selection process and incorporate e.g. AS path length or various sources of traffic engineering in the premises. On the other hand we argue that the basic inter-AS business rules and other technological constraints e.g. the role of IXP-s in the AS-AS connectivity, the multihoming opportunities or security aspects (either in the pure form of supporting secure BGP) can be rich sources of usable premises for future works.

# A

**initialization:**

1. set $r_u = (\text{acosh}(U * (\cosh(R) - 1)) + 1)$ and $\phi_u = 2\pi U$ for each node $u$;

2. sortedIDs = sort nodes according to $r_u$;

3. $\mathcal{K} = \{\text{First}(\text{sortedIDs})\}$, $E = \{\}$;

*Phase 1:*

**for** $w \to sortedIDs$ **do**
    **if** $\sum_{v \in \mathcal{K}} l(r_w, \phi_w, r_v, \phi_v) < \min_{v | r_v \leq r_w} l(r_w, \phi_w, r_v, \phi_v)$ **then**
        **forall** $v \in \mathcal{K}$ **do** $E = E \bigcup \overline{wv}$;
        $\mathcal{K} = \mathcal{K} \bigcup w$;
    **else**
        $v = \text{argmin}_{v | r_v \leq r_w} l(r_w, \phi_w, r_v, \phi_v)$;
        $E = E \bigcup \overrightarrow{wv}$;
    **end**
**end**

*Phase 2:*

**for** $\forall (u, v) : u \notin \mathcal{K} \wedge \nexists \overrightarrow{uv}$ **do**
    **if** $l(r_u, \phi_u, r_v, \phi_v) < \beta R$ **then**
        $E = E \bigcup \overline{uv}$
    **end**
**end**

Return($E$);

Figure 10: The pseudocode of YEAS